\documentstyle[12pt,twoside,psfig]{article}
\begin{document}
\begin{center}
{\Large\bf Brans-Dicke Scalar Field as a Chameleon}
\\[15mm]
Sudipta Das \footnote{E-mail:sudipta\_123@yahoo.com}~~and 
Narayan Banerjee \footnote{E-mail: narayan@iiserkol.ac.in} 
	\footnote{Present Address: IISER - Kolkata, Sector-III, 
                Salt Lake, Kolkata - 700 109, India. }

{\em Relativity and Cosmology Research Centre,\\Department of Physics, Jadavpur
University,\\ Calcutta - 700 032, ~India.}\\
[15mm]
\end{center}

\vspace{0.5cm}
\vspace{0.5cm}

\pagestyle{myheadings}
\newcommand{\be}{\begin{equation}}
\newcommand{\ee}{\end{equation}}
\newcommand{\bea}{\begin{eqnarray}}
\newcommand{\eea}{\end{eqnarray}}

\begin{abstract}
In this paper it is shown that in Brans - Dicke theory, if one considers 
a non-minimal coupling 
between the matter and the scalar field, it can give rise to a late time 
accelerated expansion for the universe preceded by a decelerated expansion for 
very high values of the Brans - Dicke parameter $\omega$. 
\end{abstract}

\section{Introduction}
During the present decade, the speculation that the universe at present 
is undergoing an accelerated phase of expansion has turned into a certainty. 
The high precision observational data regarding the luminosity - redshift 
relation of type Ia supernovae \cite{Riess}, the Cosmic Microwave Background 
Radiation (CMBR) probes \cite{Mel} suggest this 
acceleration very strongly. This is 
confirmed by the  very recent WMAP data \cite{Bridle} as well. This observation 
leads to a vigorous search for some form of matter, popularly called 
{\it dark energy}, which can drive this acceleration as normal matter cannot 
give rise to accelerated expansion due to its attractive gravitational 
properties. A large number of possible candidates for this dark energy has 
already appeared in the literature and their detailed 
behaviours are being studied 
extensively. For excellent reviews, see \cite{Sahni}. 
\par Although the expansion of the universe is accelerated at present, it 
must have had a decelerated expansion in the early phase of the evolution so as 
to accommodate for nucleosynthesis in the radiation dominated era. 
The early matter 
dominated era also must have seen a decelerated phase for the 
formation of galaxies in the universe. There are observational 
evidences too that beyond a certain value of the redshift $z$, 
the universe surely 
had a positive value for the deceleration parameter 
($q = -\frac{\ddot{a}/a}{{\dot{a}}^2/a^2} > 0$) \cite{Ag}. 
It has also been indicated 
that unless there is a signature flip from a positive to a negative 
value of $q$, the supernovae data 
are not a definite indicator of an  
accelerated expansion considering the error bars of the 
observation \cite{Tirtha}.    
\par So, we are very much in need of some form of matter, the dark energy,  
which maintained a low profile in the early part of the history of 
the universe but evolved 
to dominate the dynamics of the universe later in such a way 
that the universe smoothly 
transits from a decelerated to an accelerated phase of expansion during 
the later part of matter dominated regime. 
Apart from the Cosmological Constant $\Lambda$, 
which can indeed generate a sufficient negative pressure and 
hence drive this acceleration, 
the most talked about 
amongst the dark energy models are perhaps the `quintessence models' - a scalar 
field endowed with a potential such that the potential term evolves to  
dominate over the kinetic term in the later stages of 
evolution generating sufficient 
negative pressure which drives the acceleration. A large number 
of quintessence potentials 
have appeared in the literature (for an extensive review see \cite{Varun}). 
However, most of the quintessence potentials do not have a sound 
background from field theory explaining their genesis. Hence it might 
appear more appealing to employ a scalar 
field which is already there in the realm of the theory. 
This is where the non-minimally 
coupled scalar field models step in as the driver of this alleged late 
time acceleration.  
Brans - Dicke (BD) theory \cite{Brans} is arguably the 
most natural choice as the 
scalar tensor generalization of general relativity (GR). BD theory or its 
modifications have already proved to be useful in providing clues to the 
solutions for some of the outstanding problems in cosmology (see \cite{La} and 
\cite{NBDP}) and could generate sufficient acceleration in the matter dominated 
era even without the help of quintessence field \cite{NB}. Attempts have  
also been made to obtain a non-decelerating expansion phase for the universe 
at present by considering some interaction between the dark matter and the 
geometrical scalar field in generalised Brans - Dicke theory \cite{SDNB}. 
However, the form of interaction chosen was ad-hoc and did not follow from 
any action principle.
\par A different approach is now being considered in general relativity, where 
the quintessence scalar field is allowed to interact non-minimally with matter 
sector rather than with geometry and this interaction is introduced through an 
interference term in the action. This type of scalar field is given the name 
`chameleon field' \cite{new}. Many interesting 
possibilities with this chameleon field 
has been recently studied \cite{Khoury}. It has also been 
shown recently \cite{NSK} 
that this chameleon field can provide a very smooth transition from a 
decelerated to an accelerated phase of expansion of the universe. For 
similar work where the scalar field is strongly coupled to matter, 
see also \cite{easson}. 
However, the problem remains 
the same as that of the genesis of the scalar field. 
\par In Brans-Dicke theory or its modifications, there is an 
interaction between the 
scalar field and geometry. The chameleon field is also ``nonminimally 
coupled", but to the normal matter sector rather than with geometry. 
It deserves mention at this stage that there are attempts also 
to build up models where the dark energy and the dark matter do 
not conserve themselves individually, but has an 
interaction amongst them \cite{pavon}. One  
important motivation of considering these interactions is of 
course to seek for a 
solution of the coincidence problem - why the dark energy 
sector dominates over 
the dark matter sector now.
\par The motivation of the present work is to investigate the 
interacting models in a 
more general framework. A Brans - Dicke framework is considered, 
where there is already 
a nonminimal coupling between the scalar field and geometry. 
The action is modified to 
include a nonminimal coupling of the scalar field 
with the matter sector as well. 
This work is actually motivated 
by the recent work by Clifton and Barrow \cite{Clifton} where they studied 
the behaviour of an isotropic cosmological model in the early as well as 
in the late time limits in this framework. However, the nonminimal 
coupling of a scalar field with both of geometry and the 
matter sector has been in use for quite a long time, 
courtesy the dilaton gravity, the low energy limit of string theory. 
The present work uses the ansatz for a particular purpose, namely to check 
if the required signature flip 
in the deceleration parameter $q$ can be obtained from this model. 
The actual form of the  
coupling of the scalar field with matter certainly has to be 
introduced by hand, but the 
model has the advantage of having the scalar field in the theory itself.
\par As already mentioned, although Brans-Dicke theory proved useful 
for the solution for many a 
cosmological problem, it has the serious drawback that the 
Brans-Dicke parameter $\omega$ 
has to have a small value of order unity. This squarely 
contradicts the local astronomical 
requirement of a pretty high value of $\omega$. It has been shown that the 
present model works even for very
high values of $\omega$ ($\sim 10^4$) and thus can have good 
agreement with the observational limits \cite{Iess, Dicke}. 
Hence this kind of general interaction has features, 
which might solve the cosmological problems as well as 
take care of the observations on 
the solar systems etc.

\par In the next section the model is described and it is shown that this 
type of non-minimally coupled interacting models can provide a smooth 
transition from decelerated to accelerated phase of expansion for a wide 
range of values of the BD parameter $\omega$. Section 3 gives 
two cases of exact solutions and the last section discusses the results. 
\section{Field Equations and Results :~}
The relevant action in BD theory is given by 
\be\label{action}
A = \int{\sqrt{-g}}~d^4x\left[\frac{\phi R}{16 \pi G} + \frac{\omega}{\phi}
               {\phi}^{,\mu}{\phi}_{,\mu} + L_{m}f(\phi)\right]~, 
\ee 
where $R$ is the Ricci scalar, $G$ is the Newtonian constant of gravitation, 
$\phi$ is the BD scalar field which is non-minimally coupled to gravity, 
$\omega$ is the dimensionless BD parameter.  
The last term in the action indicates the interaction between the matter 
Lagrangian $L_{m}$ and some arbitrary function $f(\phi)$ of the BD scalar 
field. If $f(\phi) =$ constant $= 1$, one gets back the usual BD action. 
\\
For a spatially flat FRW model of the universe, the line element is given by 
\be 
ds^2 = dt^2 - a^2(t) \left[ dr^2 + r^2d{\theta}^2 + r^2\sin^2{\theta} 
                      d{\phi}^2 \right]~,
\ee
where $a(t)$ is the scale factor of the universe. 
\\
Variation of the action (\ref{action}) with respect to the metric components 
yields the field equations as 
\be\label{feq1}
3\frac{\dot{a}^2}{a^2} = \frac{{\rho}_m f}{\phi} + \frac{\omega}{2}
    \frac{\dot{\phi}^2}{\phi^2} - 3 \frac{\dot{a}}{a}\frac{\dot{\phi}}{\phi}~,
\ee
\be\label{feq2}
2\frac{\ddot{a}}{a} + \frac{\dot{a}^2}{a^2} = -\frac{\omega}{2}
      \frac{\dot{\phi}^2}{\phi^2} - \frac{\ddot{\phi}}{\phi} 
              - 2\frac{\dot{a}}{a}\frac{\dot{\phi}}{\phi}~.
\ee
Here $\rho_{m}$ is the energy density of dark matter and as 
the universe at present is dominated by matter, the fluid is taken in 
the form of pressureless dust, i.e, $p_{m} = 0$. Here, a dot indicates 
differentiation with respect to the cosmic time $t$.\\
Also, variation of action (\ref{action}) with respect to the Brans - Dicke 
scalar field $\phi$ yields the wave equation as 
\be\label{waveeq} 
(2\omega + 3) \left(\ddot{\phi} + 3 \frac{\dot{a}}{a}\dot{\phi}\right)  = 
               \rho_{m}f + \rho_{m} f' \phi~, 
\ee
where a prime indicates differentiation with respect to $\phi$.
\\ From the two field equations and the wave equation one can 
arrive at the matter conservation equation which comes out as 
\be\label{matter} 
\dot{\rho_{m}} + 3 \frac{\dot{a}}{a}\rho_{m} = 
                  -\frac{3}{2}\rho_{m} \frac{\dot{f}}{f}
\ee
and readily integrates to yield 
\be\label{matterconsrv}
\rho_{m} = \frac{\rho_{0}}{a^3 f^{3/2}}~,
\ee
where $\rho_{0}$ is a constant of integration. \\
It is evident from equation (\ref{matterconsrv}) that the usual matter 
conservation equation gets modified here because the scalar field 
is now coupled to both geometry and matter.  
\par Out of equations (\ref{feq1}), (\ref{feq2}), (\ref{waveeq}) and 
(\ref{matterconsrv}), only three are independent equations as the fourth 
one can be derived from the other three in view of the Bianchi identity. On 
the other hand, we have four unknowns - $a$, $\rho_{m}$, $f(\phi)$ and $\phi$ 
to solve for. 
\\
In order to close the system of equations, we make an ansatz
\be\label{alpha}
\frac{\dot{\phi}}{\phi} = -\frac{\alpha}{H}~,
\ee
where $\alpha$ is an arbitrary positive constant. There is no a priori 
physical motivation for this choice, this is purely phenomenological which 
leads to the desired behaviour of the deceleration parameter $q$ of 
attaining a negative value at the present epoch from a positive value 
during a recent past. The system of equations 
is closed now. Some characteristics of the model can now be discussed even 
without solving the system. In the remaining part of the section, 
the possibility of having a transition of the mode of the expansion 
from a decelerated to an accelerated one is studied. \\
With the assumption (\ref{alpha}), 
equation (\ref{feq2}) easily yields an expression for 
the deceleration parameter $q$ as 
\be\label{eqq}
q = \frac{H^2 +\left(\frac{\omega}{2}+1\right)\frac{\alpha^2}{H^2}-3 \alpha}
           {2 H^2 + \alpha}~.
\ee
In obtaining equation (\ref{eqq}), the relation 
\be\label{dotH} 
\dot{H} = -H^2 (q + 1)
\ee 
has been used. $H$ gradually decreases with time from a very large value at 
the beginning of the evolution. The equation (\ref{eqq}) indicates that 
$q = \frac{1}{2}$ at $H \rightarrow \infty$, i.e, the model starts exactly 
the same way as a matter dominated spatially flat model does. 
\par Now the deceleration parameter $q$ is plotted against $H$ ( Figure 1(a)
- 1(h) ) for
different values of $\omega$ and $\alpha$. They clearly show 
that the required signature flip in $q$ can be
obtained for any negative value of $\omega$ and also for small positive
values of $\omega$ in some recent past ($H > 1$). The nature of the behaviour 
of $q$ against $H$ is hardly affected by a small change in the value of 
$\omega$, which can only shift the epoch at which the acceleration sets in. 
This can again be adjusted by choosing the value of $\alpha$ properly which is
a free parameter. It must be mentioned that as $\frac{1}{H}$
is a measure of the age of the universe and $H$ is a monotonically decreasing
function of time $t$, `future' is given by $H < 1$, `past' by $H > 1$ if
$H$ is scaled by the present value $H_{0}$, i.e, $H = 1$ at the present
epoch. \\

\begin{figure}[!h]
\centerline{\psfig{figure=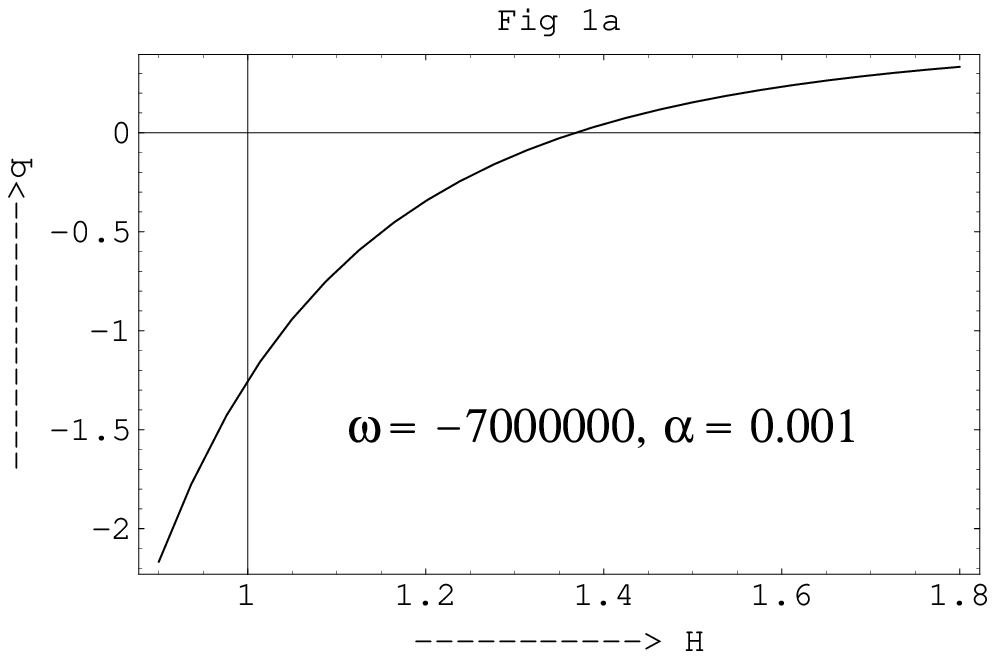,height=50mm,width=80mm} 
\psfig{figure=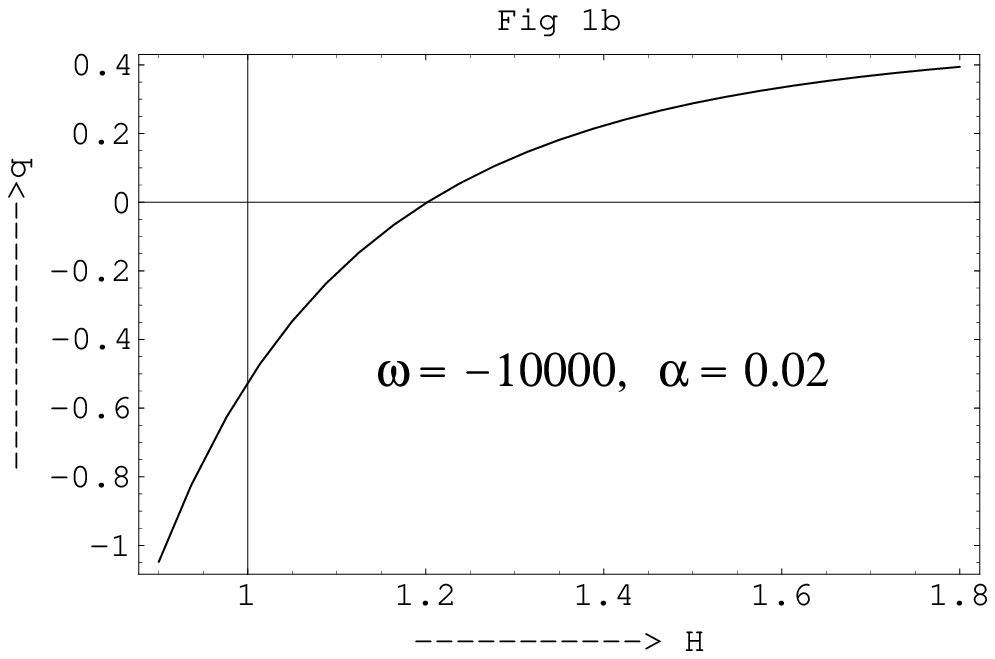,height=50mm,width=80mm}}
\end{figure}
\begin{figure}[!h]
\centerline{\psfig{figure=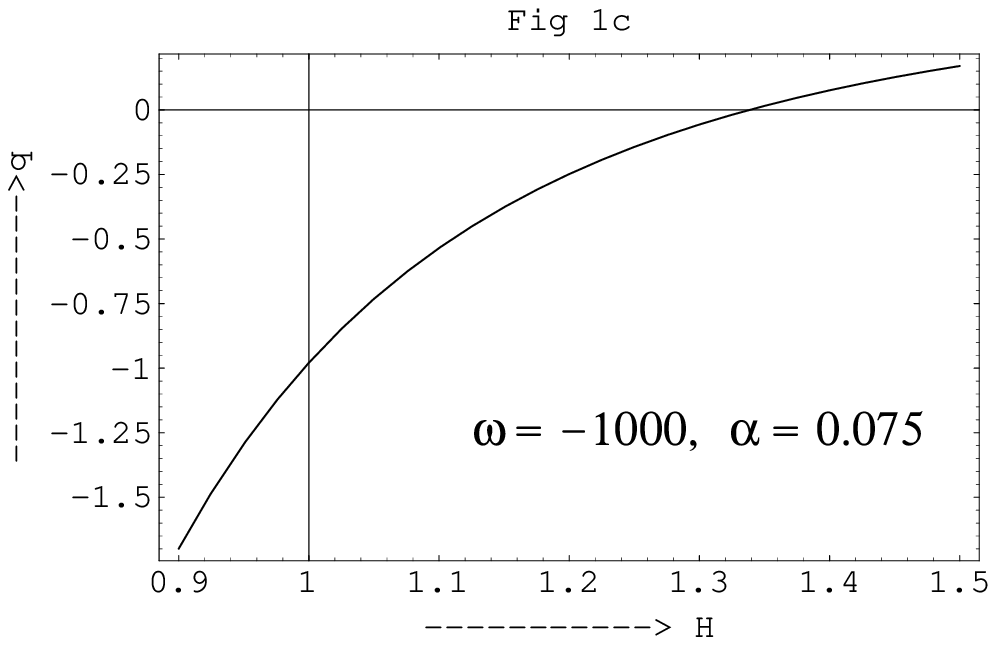,height=50mm,width=80mm} 
             \psfig{figure=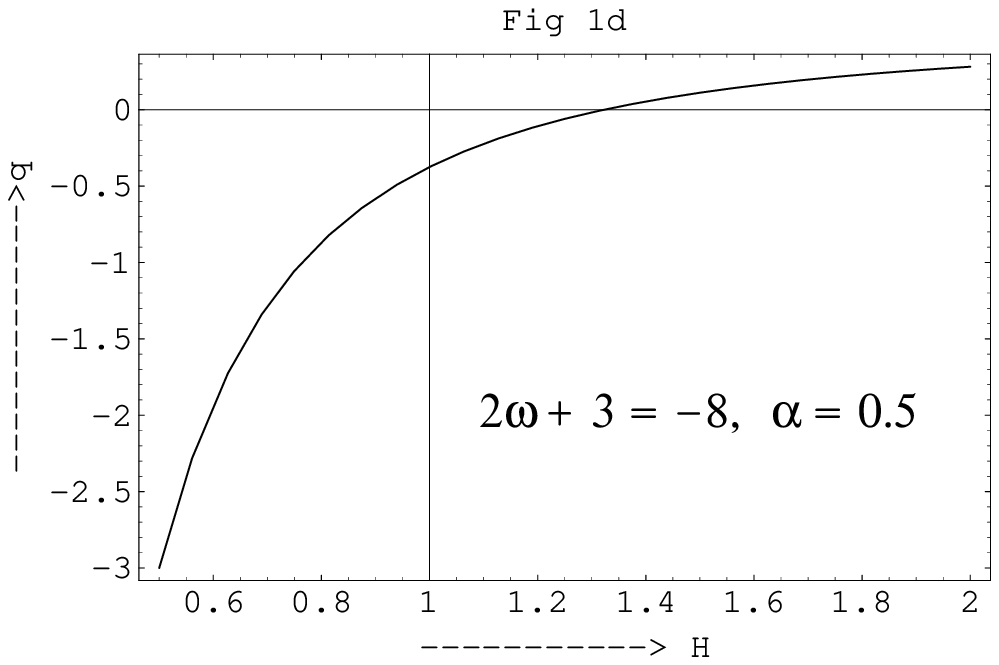,height=50mm,width=80mm}}
\end{figure}
\begin{figure}[!h] 
\centerline{\psfig{figure=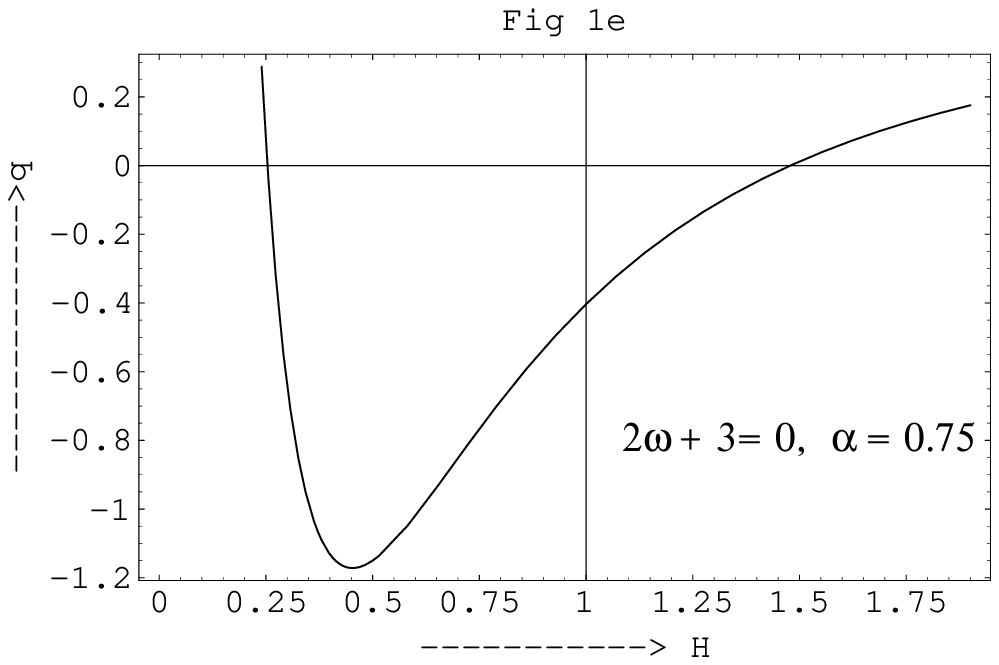,height=50mm,width=80mm}
\psfig{figure=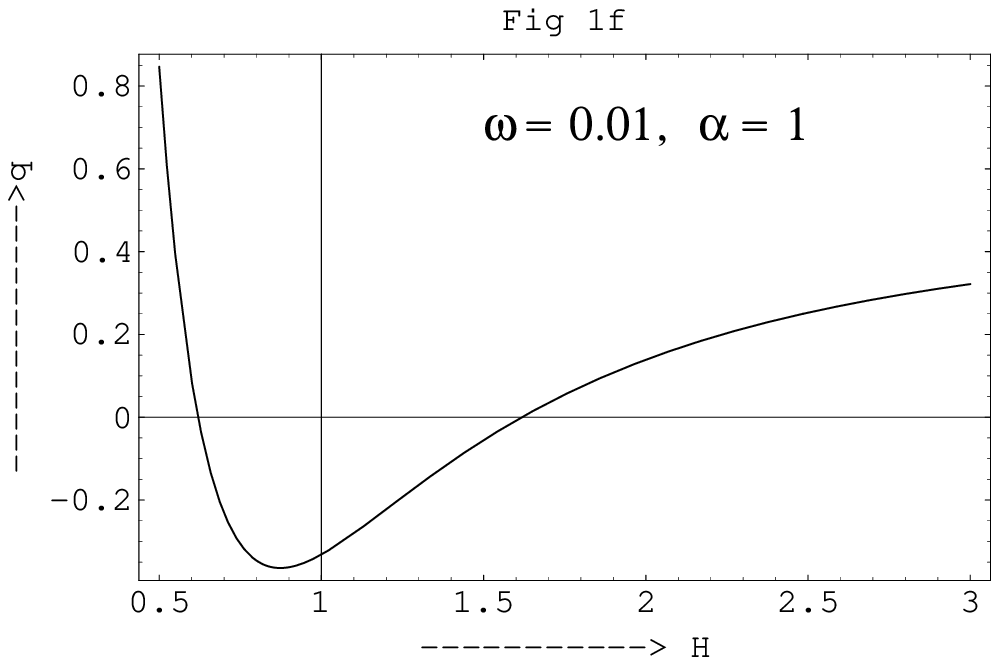,height=50mm,width=80mm}}
\end{figure}
\begin{figure}[!h] 
\centerline{\psfig{figure=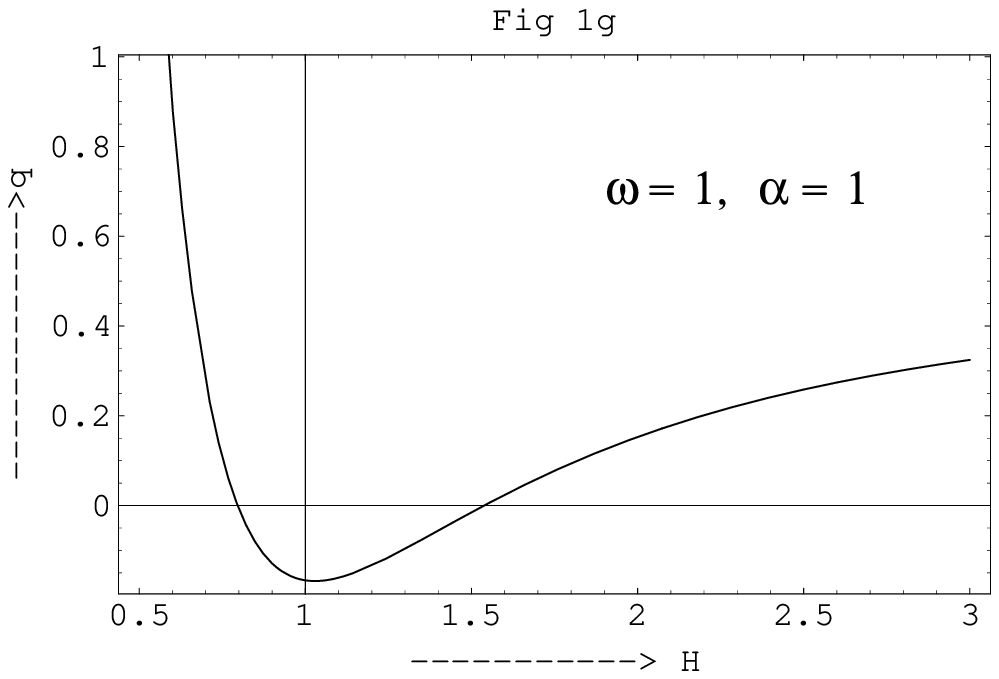,height=50mm,width=80mm}
\psfig{figure=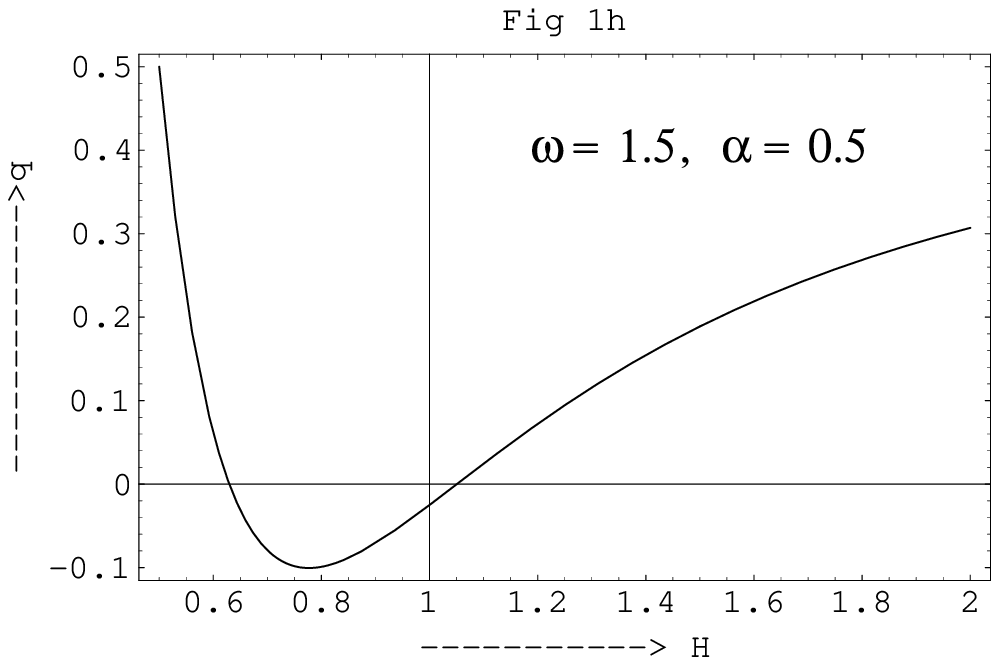,height=50mm,width=80mm}}
\caption{\normalsize{The $q$ vs. $H$ plot for different values of $\omega$ 
and $\alpha$}}
\label{qvsh}
\end{figure}

Figures 1(e) - 1(h)  have the additional feature that $q$ has two 
signature flips. For example, in figure 1(e) (i.e, for $\omega = 
-\frac{3}{2}$), the flips take place around $H \approx 1.5$ (i.e., past) and  
$H \approx 0.25$ (future). So, in all these cases the universe reenters a 
decelerated phase of expansion again in near future and thus a `phantom menace' 
is avoided, i.e, the universe does not show a singularity of infinite 
volume and infinite rate of expansion in a `finite future'. The cases 
1(a) - 1(d), however, do not have this ``double signature flip'', which 
can be seen from equation (\ref{eqq}). If $q$ is put equal to zero, the 
combinations of values of $\omega$ and $\alpha$ used in these figures 
do not yield two real positive roots for $H$. 

\section{Two specific examples:}
Equation (\ref{alpha}) along with equation (\ref{dotH}) can be written as 
\be
\frac{d}{dH}\left(\ln \phi\right) = \frac{\alpha}{H^3 (q + 1)}~.
\ee
Replacing the expression for $q$ from equation (\ref{eqq}), the above equation 
takes the form 
\be\label{eqnphi}
\frac{d}{dH}\left(\ln \phi\right) = \frac{\alpha\left(2 H^2 + \alpha\right)}
              {3 H\left[H^4 - \frac{2 \alpha}{3}H^2 + \frac{1}{3}
                   \left(\frac{\omega}{2} + 1\right)\alpha^2\right]}~.
\ee
We try to solve equation (\ref{eqnphi}) analytically for two special cases.\\
\\
{\bf{Case I~: $2\omega + 3 = 0$}}\\
\par With this choice of $\omega$, equation (\ref{waveeq}) immediately gives
\be\label{omega1}
f = \frac{\phi_{0}}{\phi}~,
\ee
$\phi_{0}$ being a constant of integration. \\
Equation (\ref{eqnphi}) then leads to the solution for $\phi$ as 
\be
\phi = A~ \frac{{H^4\left(H^2 - \frac{\alpha}{2}\right)}^2}
           {{\left(H^2 - \frac{\alpha}{6}\right)}^4}~,
\ee
$A$ being a constant of integration. \\
Using this expression for $\phi$, one can obtain the solutions for $a$, 
$\rho_{m}$ and $H$ as 
\be
a = a_{0} \frac{{\left(H^2 - \frac{\alpha}{6}\right)}^{\frac{2}{3}}}
 {{\left[\left(H^2 - \frac{\alpha}{2}\right)\left(3H^4 - 3 \alpha H^2 + 
\frac{3 \alpha^2}{4}\right)\right]}^{\frac{1}{3}}}~,
\ee
\be
\rho_{m} = \frac{3 A^2}{\phi_{0}}\frac{ H^6{\left(H^2 - \frac{\alpha}{2}
 \right)}^6}{{\left(H^2 - \frac{\alpha}{6}\right)}^8}~~~~~~~~~~
\ee
\be 
\textrm{and}~~\left(\frac{H - \sqrt{\frac{\alpha}{2}}}
  {H + \sqrt{\frac{\alpha}{2}}}\right) \left(\frac{H + \sqrt{\frac{\alpha}{6}}}
  {H - \sqrt{\frac{\alpha}{6}}}\right)^{\frac{2}{\sqrt{3}}} = 
        \exp\left(\sqrt{\frac{\alpha}{2}}\left(t_{0} - t\right)\right)~.
\ee
\\
It deserves mention that $\omega = -\frac{3}{2}$ is a special case because 
in conformally transformed version of the BD theory, $(2\omega + 3) = 0$ 
indicates that the kinetic part of the energy contribution from the scalar 
field sector is exactly zero \cite{Dicke}.  
\par As shown in equation (\ref{omega1}), this 
particular choice of $\omega$ gives $f(\phi)\sim \frac{1}{\phi}$. The 
converse is also true. If one starts by assuming $f(\phi)\sim \frac{1}{\phi}$, 
$\omega$ can take only one value, i.e., $-\frac{3}{2}$ and one arrives 
at the same results. 
\par As already mentioned this choice of $\omega = -\frac{3}{2}$ provides 
the important feature of `future' deceleration and thus does not suffer 
from the problem of `big rip'.
\\
\\
{\bf{Case II~: $2\omega + 3 = -8$}}\\
In this case, equation (\ref{eqnphi}) yields the solution for $\phi$ as 
\be 
\phi = \phi_{0}\left( 1 - \frac{7 \alpha}{6 H^2}\right)^{\frac{2}{7}}~,
\ee
$\phi_{0}$ being a constant of integration. \\
Using this expression for $\phi$, from equation (\ref{alpha}) one can obtain 
the solutions for the Hubble parameter $H$ and the scale factor $a$ as 
\be
H = \sqrt{\frac{7 \alpha}{6}} \coth \left(\frac{3}{2}
                   \sqrt{\frac{7 \alpha}{6}}~t\right)~,
\ee
\be\label{a2}
a = a_{0}\left[\sinh \left(\frac{3}{2}\sqrt{\frac{7 \alpha}{6}}~t\right)
               \right]^{\frac{2}{3}}~.
\ee
The interesting feature of equation (\ref{a2}) is that for small $t$, $a \sim 
t^{\frac{2}{3}}$ which is same as that for a dust dominated era. On the other 
hand, for high values of $t$, $a \sim e^{\frac{3}{2}
\sqrt{\frac{7\alpha}{6}}t}$ and thus gives an accelerated expansion for the 
universe. \\
From the field equations, the solutions for $f$ and $\rho_{m}$ also comes 
out as 
\be\label{fII}
f(\phi) = \frac{196 {\rho_{0}}^2}{{a_{0}}^6 \alpha^2} \frac{1}{\phi^2}\left[
   \frac{1}{40{\left(\frac{\phi}{\phi_{0}}\right)}^{-\frac{7}{2}} - 24 + 
   33{\left(\frac{\phi}{\phi_{0}}\right)}^{\frac{7}{2}}}\right]^2
\ee
and 
\be
\rho_{m}(t) = \frac{{a_{0}}^6 {\alpha}^3 {\phi_{0}}^3}{2744{\rho_{0}}^2}
   \left(\textrm{sech}^{\frac{12}{7}}X\right) \left[40~\textrm{coth}^2 X - 
24~\textrm{cosech}^2 X + 
         33~ \textrm{sech}^2 X ~\textrm{cosech}^2 X \right]
\ee
where $X = \frac{3}{2}\sqrt{\frac{7 \alpha}{6}}~t$.
\par Although here the equation system has been completely solved for only 
 small negative values of $\omega$, this model works even for high values of 
$\omega$ ( $\sim 10^5$ ) as shown in figure 1(a) - 1(c). This is consistent 
with the limits imposed by solar system experiments which predict the 
value of $\omega$ to be of the order of tens of thousands 
( $\vline~{\omega}~\vline \ge 40000$ ) \cite{Iess}. 
\par It deserves mention that in figure 1(a) or 1(b), where values of 
$\omega$ chosen are very high ($\omega = 10^6$ in fig 1(a) and $\omega 
= 10^4$ in fig 1(b)), the corresponding values of $\alpha$ 
required are very low ($\alpha = 
10^{-3}$ in fig 1(a) and $\alpha = 0.02$ in fig 1(b)) in order to adjust  
the time of signature flip in observationally consistent region.  
\section{Discussion :}
Thus we see that for a spatially flat FRW universe ($k = 0$), we can construct 
a presently accelerating model with the history of a deceleration in the past in Brans - Dicke 
theory by considering a coupling between the matter Lagrangian and the 
geometric scalar field. The salient feature of this model is that no dark 
energy sector is required here to drive this alleged acceleration. 
Also it deserves mention that the nature of the $q$ vs. $H$ plot is not 
crucially sensitive to the value of $\omega$ chosen; only the `{\it time}' 
when the signature flip in $q$ occurs shifts a little but that too can be 
adjusted by properly choosing the value of $\alpha$, which is a parameter of 
the model. 
\par The matter conservation equation obviously gets 
modified in this framework due to the coupling 
between matter and the scalar field, i.e., 
matter is no longer conserved by itself.  
The right hand side of equation (\ref{matter}) indicates that 
a transfer of energy between the 
matter and the scalar field takes place due to the coupling factor $f(\phi)$. 
One may have an idea about the direction and amount of this energy transfer 
if $f(\phi)$ is exactly known. In the two specific examples discussed in the 
present work, the energy infact flows from the dark matter to the scalar 
field sector. In case I, with $2\omega + 3 = 0$, $f(\phi)$ is given by 
equation (\ref{omega1}) which yields ( with equation (\ref{alpha}) ) 
\begin{center}
$\frac{\dot{f}}{f} = -\frac{\dot{\phi}}{\phi} = + \frac{\alpha}{H}~.$
\end{center}
So the right hand side of equation (\ref{matter}) is negative and hence 
$\rho_{m}$ decreases more rapidly than what is expected for a self - conserved 
matter sector. As we are working in units where the present value of $H$ is 
equal to 1 and $\alpha$ is less than one ( $\alpha = 0.75$ as used in 
figure 1(e) ), the present transfer rate is obviously less than the Hubble 
rate of expansion. In case II, where $2\omega + 3 = -8$, one can use 
equations (\ref{fII}) and (\ref{matter}) to find $\frac{\dot{f}}{f}$, and 
if $\alpha < 1$, the transfer rate is of the same order of magnitude 
as the Hubble expansion rate. In this case also, the present 
$\frac{\dot{f}}{f}$ is positive and hence the energy flows from the 
dark matter sector to the scalar field sector. If 
$f(\phi) = \textrm{constant}$, this interaction vanishes and the matter sector 
conserves itself as usual. 
\par As the nonminimally coupled scalar field theories allow for a 
variation of the strength of gravitational interaction, it is worthwhile 
to comment on this aspect as well. As $\frac{1}{\phi}$ behaves as the 
effective Newtonian gravitational constant $G$, one has 
\begin{center}
$\frac{\dot{G}}{G} = -\frac{\dot{\phi}}{\phi} = + \frac{\alpha}{H}~.$
\end{center}
As the present value of $H = 1$, and $\alpha \le 1$ in all the examples 
discussed, $\frac{\dot{G}}{G}$ at present is less than the Hubble rate 
of expansion. As already mentioned, the signature flip in $q$ in the 
figures 1(a) to 1(h) can still be obtained with other choices of the 
pair of $\omega$ and $\alpha$, the value of $\frac{\dot{G}}{G}$ can 
be further lowered. In the early stages, when $H$ had a very high value, 
$\frac{\dot{G}}{G}$ was infact negligible. In far future when 
$H \rightarrow 0$, $\frac{\dot{G}}{G}$ may have high values, but at that 
epoch the hierarchy between gravitational and electroweak couplings will 
hardly matter. 
\par As mentioned earlier, this particular model works for a wide range of 
values of $\omega$ and even for high values of $\omega$ 
($\sim 10^4$). So this model is capable 
of solving two major problems at one go - the first one is to obtain 
the smooth transition from a decelerated to an accelerated phase of 
expansion in the recent past without any dark energy sector 
and the second one is to solve the nagging 
problem of discrepancy in the values of $\omega$ as suggested by local 
experiments and that required in the cosmological context. It had been shown 
before that Brans-Dicke scalar field interacting with dark matter can indeed 
generate an acceleration \cite{SDNB} where $\omega$ is not severely restricted 
to low values, but the parameter $\omega$ had to be taken as a function of the 
scalar field $\phi$. The Brans-Dicke scalar field interacting nonminimally 
with dark energy sector also has a possibility of having an arbitrary value of 
$\omega$ \cite{MPLA}. But again that required a dark energy sector 
as the driver of the acceleration.
\par It deserves mention at this stage that the belief 
that BD theory goes over to GR in 
the high $\omega$ limit suffered a jolt \cite{nbss}. But in the weak 
field regime, relevant for the observations on the solar system, 
a high value of 
$\omega$ is still warranted \cite{Iess, Dicke}. So the present work, 
and the work by 
Clifton and Barrow \cite{Clifton} indeed opens up the 
possibility of seeking solutions 
to cosmological problems in BD theory. It should be noted  
that in view of the coupling between $\phi$ and $\L_{m}$ as $f(\phi)L_{m}$ 
in the action, it is required that the said weak field approximation of the 
field equations be re-visited. In the presence of a potential $V = V(\phi)$ 
where $\phi$ is the BD field, such investigations have already been 
there \cite{olmo}, where the results depend on derivatives of the potential. 
For the present work, the details of the calculations will be different as 
there is no potential $V(\phi)$ as such. Investigations in this direction 
to find the actual order of magnitude of $\omega$ which passes the 
astronomical `{\it fitness test}' are in progress. However, it appears that 
although the expression for the post - Newtonian corrections for various 
modifications of BD theory will have different features, all will 
require a large value of $\omega$ for the local astronomical tests 
\cite{olmo, tclifton}. 
\par In the context of the present accelerated expansion of the universe, 
non-linear contribution from the Ricci scalar $R$ in the action has 
attracted a lot of interest \cite{fr}. This form of action, very widely 
dubbed as $f(R)$ gravity, has been shown to be formally equivalent to 
a Brans - Dicke action endowed with an additional potential $V$ which 
is a function of the BD scalar field for a particular value of the BD 
parameter $\omega$, namely $\omega = -3/2$ \cite{vollick}. 
This is particularly true for the Palatini kind of variation for the 
$f(R)$ gravity action. The present work does not contain a $V = V(\phi)$, 
but the term $f(\phi)L_{m}$ in the action serves as an effective potential 
and may serve a similar purpose. 
\par The particular interaction 
chosen in this work is contrived, but it at least serves as a 
toy model, where the very existence of the particular scalar field 
is not questioned, it is there in the theory. A form 
of $f(\phi)$ for which the dark matter sector redshifts 
close to $a^{-3}$, and the acceleration takes place for 
quite a high value of $\omega$, could indeed be a very interesting 
possibility. Furthermore, the model has a lot of features, and has the 
promise to reproduce other forms of modifications of gravity as special 
cases.

\section{Acknowledgement }
One of the authors (SD) wishes to thank CSIR for financial support. 

\end{document}